\documentclass[11pt,twocolumn]{article}

\usepackage{abstract}
\usepackage{graphicx}
\usepackage{float}

\newcommand{\mind}{{\small {MMC}}\,}
\newcommand{\mmc}{{\small MMC}}

\newcommand{\LCDM}{{\small {$\Lambda$CDM}}\,}
\newcommand{\asympt}{\sim}
\newcommand{\f}{$f$}
\newcommand{\fb}{$f_b$}
\newcommand{\m}{$m$}
\newcommand{\M}{$M$}
\newcommand{\w}{$w$}
\newcommand{\G}{$G$}

\newcommand{\wurzq}{\mbox{$1-\frac{v^2}{c^2}$}}
\newcommand{\wurz}{\mbox{$\sqrt{1-\frac{v^2}{c^2}}$}}
\newcommand{\rhoinf}{\mbox{$\rho_{\infty}$}}

\newcommand{\be}{\begin{equation}}
\newcommand{\ee}{\end{equation}}
\newcommand{\benum}{\begin{enumerate}}
\newcommand{\ds}{\displaystyle}
\newcommand{\eenum}{\end{enumerate}}

\newcommand{\vp}{\mbox{$\vec{p}$}}
\newcommand{\vv}{\mbox{$\vec{v}$}}
\newcommand{\vr}{\mbox{$\vec{r}$}}
\newcommand{\vvp}{\mbox{$\dot{\vec{v}}$}}

\renewcommand{\theta}{\vartheta}
\title{
A Modified Mass Concept could explain both the Dark Matter and the Dark Energy Phenomenon
}

\author{ \sc{Willi R. B\"ohm}\\
         \it{Fachhochschule W\"urzburg--Schweinfurt, Germany}\\
         \it{University of Applied Sciences}\\
         \tt{\small e-mail: Willi.Boehm@fhws.de}}

\date{}
\begin{document}

\twocolumn[
\maketitle
\begin{onecolabstract}
% version 0.99 5.12.2010
Some consequences of a Modified Mass Concept (MMC) are discussed.
According to MMC the inertial mass is not only determined by its
energy, but also by a scalar field f depending on other masses.
The concept consistently describes the galactic rotation curves,
the inflation of the universe and its accelerated expansion, all
without the necessity of Dark Matter (DM) and Dark Energy (DE).
Instead, the effects attributed to DM are caused by a reduction of
inertia acting as an enhancement of gravity. These results of MMC
are similar to that of MOND. The effects usually attributed to DE
in MMC stem from a new equation of state for baryonic matter,
which always causes a negative pressure. In this respect the
results are similar to those of the two component LCDM. In
particular, according to MMC the late universe will pass into a
state of constant energy density. Furthermore, the MMC can provide
an explanation of the very high peculiar velocities found at large
scales.

\end{onecolabstract}]

%\onecolumn

% version 0.99 5.12.10
\section{Introduction}
\label{Sek1}

The "missing mass problem" at large scales is well-known in astronomy since Zwicky \cite{Zwicky}. It applies to
galaxies, clusters und the universe as a whole. In galaxies and clusters the observed mass cannot explain the observed
dynamic. The amount of observed mass is always smaller than that of the calculated virial mass. The well-established solution for this problem is the
postulation that there exists Dark Matter interacting with baryonic matter only by
gravitation \cite{Gentile04,Biviano06,Kassin06}. In the meantime, some concerns have been raised  over the paradigm
of DM \cite{McGaugh04,McGaugh09,Famaey05,Kroupa10}.

 A popular alternative theory is the MOdified Newton Dynamic (MOND) developed by M. Milgrom. Today there exists a lot of
MONDian literature. A non-representative selection is e.g. \cite{Milgrom93,Milgrom98,Milgrom99,Milgrom01,Milgrom02,Milgrom05}
and \cite{Bekenstein84}. An extensive and actual list of MOND literature is in \cite{McGaugh}. MOND is very successful with
respect to the galactic rotation curves \cite{Sanders96}, galaxy formation \cite{Sanders07} and the simulation of the evolution
of spirals \cite{Tiret07}. But in the past, there are also problems reported amongst others with lensing \cite{Hoekstra02},
with temperature profils of cluster \cite{Aguirre01}, with the Ly$ \alpha$ forest \cite{Aguirre01a} and the ominous
"bullet cluster"\footnote{an interesting discussion about this issue see \cite{McGaugh}}.\\

The "missing mass problem" of cosmology results from the observation that the density of the visible matter together with
the DM is too small to cause the flatness of space, which is an established fact today.
Furthermore, there is a need to explain why the universe had an accelerated expansion in the recent past (z= 0.5 \ldots 1).
This mostly is attributed to a special form of energy, called Dark Energy (DE). A good review on this topic is \cite{Copeland06,Sami09}.
According to the current conception, the universe today consists of about 70\% DE and baryonic matter and DM add up to
about 30\%. The most simple variant of DE is the cosmological constant $\Lambda$ \cite{Carroll01}.
The nowadays standard cosmology is the \LCDM\, which combines the cosmological constant and the cold dark matter.\\

The presented work exhibits some parallels to both the MOND and the two component \LCDM.\\

Section \ref{basicidea} deals with the basic idea of \mmc. The issue of section \ref{mmcNewton} is the non relativistic
equation of motion and the difference between MOND and \mmc. Section \ref{mmcrelmech} is about the relativistic equation
of motion in SR and GR. The new matter equation of state in \mmc\, is deduced in section \ref{mmccosmol}.
Finally, in section \ref{observ} the predictions of \mmc\, are compared to the observational facts.

% version 099  5.12.10
\section{The Basic Idea}\
\label{basicidea}

In \mmc\, the basic assumption is made  that inertia of a mass \m\, depends not only on its localized energy content,
but there is also a non-local effect caused by a scalar field \f, which another distant mass generates.\\

In the non-relativistic case of a test mass \m\, at \vr\, in the gravitational field of a mass \M\, causing the field
\f\, we may write:\\
\be
\label{1}
 m_{in}=f(\vr) m_{g}
\ee
with the gravitational mass $m_{g}$ and the inertial mass $m_{in}$ of \m\, with respect to \M.\\

At first glance this seems to be the revival of a very out-dated debate. Didn't the famous E\"otv\"os experiment
\cite{Eotvos1, Eotvos2} and its successors \cite{Dicke,Braginsky} and more recently \cite{Schlamminger} show  that
$m_g$ {\it is equal to} $m_{in}$ with very high precision, as we all have internalized? If we reconsider the exact
conclusion of an E\"otv\"os balance experiment, we realize that it has  only shown the precise proportionality
of the both quantities, and any multiplicative constant can be put into the definition of force. Thus, according to
eq.(\ref{1}) any E\"otv\"os-like experiment fixed on earth will deliver a zero result in \mmc\, too. However, \mmc-effects
could be detected in a satellite based experiment in which \f\, changes\footnote{Unfortunately,  the STEP mission \cite{STEP},
a test for the equivalence principle, will be flown in a near-circular orbit for different reasons. We may have to
wait for LISA \cite{LISA} the Laser Interferometer Space Antenna of NASA and ESA, designed for the detection
of gravitational waves.}.\\

According to eq.(\ref{1}), \f\, is dimensionless. The absolute strength of \f\, will be not fixed here because it
will be beneficial to scale \f\, in such a way that $f \approx 1$ holds in the vicinity of a large mass.
Then the absolute value of \f\, can be merged with the gravitational constant \G. Of course that would mean a rescaling
of the gravitational strength.\\

Requirements on the scalar field \f:
\benum
\item[r1:] $0 \le f \le 1$ because of scaling.
\item[r2:] $f \approx 1$ in the proximity of a large mass \M.
\item[r3:] there exists a scale length $\lambda$ depending on \M.
\item[r4:] \f\, decreases monotonically with the distance from \M\, and at large distances $r$, $ f \asympt \lambda/r$
           asymptotically holds.
\eenum

The scale length $\lambda$ is that distance from \M\, at which the gravitational acceleration is $a_C$, thus
$\lambda=\sqrt{\frac{G M}{a_C}}$. Therein is $a_C=\alpha c H_0$, $c$ the speed of light, $H_0$ the Hubble constant of the
present epoch and $\alpha$ is a fit parameter.\footnote{
$\alpha=0.169$ yields the MOND-Parameter $a_M$(see below)}${}^,$\footnote{in the cosmological context $\lambda$
depends on the asymptotic mass density (see below)}\\

In Special Relativity (SR), we have $m_g=m$ and $m_{in}=f m_{rel}$ with the rest mass $m$ and $m_{rel}=\frac{m}{\wurz}$.
In General Relativity (GR) the distinction between $m_{g}$ and $m_{in}$ is artificial, because the energy-momentum-tensor
determines the metric. Thus, in GR we have de facto
$ m_{g}=m_{in}=f m_{rel} $.\\

Because  mass plays a paramount role in all of physics, this basic assumption has momentous consequences. In this article I
restrict myself to a few issues with direct impact on astronomy and cosmology. At the moment there is no field
function \f\, deduced from first principles. Therefore, in sections \ref{basicidea} to  \ref{mmccosmol} consequences
are discussed which do not depend on the special form of \f. When comparing \mmc\, with the observation in
section \ref{observ} empirical functions will be used.\\

In this article the symbols have their usual meanings. For example, the momentum of a mass \m\, with velocity $\vv$ in \mmc\,
is
\be
\label{2}
 m_{in}\vv=f\vp
\ee
with $\vp=m_{rel} \vv$.

%version 0.99   5.12.10

\section{\mmc\, and Newtonian Mechanics}
\label{mmcNewton}
\subsection{Consequences of the Equation of Motion}

The equation of motion of a test mass \m\, in a Newtonian potential $\Phi_N$ of a large mass \M\, is given by the
second Newtonian law
\be
\label{3}
\frac{d}{dt}\left( f(\vr(t)) m_{g} \vv \right) = -m_{g} \nabla \Phi_{N}
\ee
or
\be
\label{3a}
 f m_{g} \dot{\vv} +  \dot{f} m_{g} \vv = -m_{g} \nabla \Phi_{N}
\ee

Applying d'Alembert's principle we realize that  generally there are at least two inertial forces, even in a linear motion:
the well-known inertial force $ -f m_g \dot{\vv}$ resisting a change of velocity and additionally $-\dot{f} m_g  \vv$
favoring the decrease of inertial mass, because this force is always repulsive. If \m\, approaches \M\ then $\dot{f}>0$,
whereas with increasing distance  $\dot{f}<0$ holds. In the cosmological context an analogue term will cause a negative
pressure.\\

It is useful to solve eq.(\ref{3a}) for the acceleration
\be
\label{4}
 \dot{\vv}=\frac{1}{f}(-\nabla \Phi_N -\dot{f} \vv )
\ee
which gives  the standard physics results for $f=1$.

One can see that the weak equivalence principle is valid, because the trajectory does not depend on $m_{g}$.
Furthermore, we learn from eq.(\ref{3}) that the momentum $f \vp$ is conserved if $\nabla \Phi_N=\vec{0}$.\\

In the strict sense the strong equivalence principle holds only approximately if $f \approx const$, for example
close to a large mass. Because the principle has to be valid only locally it is a good approximation especially in
cosmology, where \f \,changes only on very large scales.\\

Now let us discuss some consequences arising from the gravitational term.\\

Since, by assumption $f \le 1$, the factor $1/f$ acts as an enhancement of gravity. In \mmc\, a free falling body is
accelerated by a factor $1/f$  more than in standard theory. This works like an additional gravitational mass.
One can  easily imagine that the enhancement factor could  tremendously effect the development of structures at
large scales. This applies to the formation of galactical spiral arms as well as to the emergence of voids and filaments
at the level of super clusters.\\

In the end it accounts also for the observed galactical rotation curves. To show the flatness of the rotation curves in
the outskirts of galaxies, we examine the special case of a test mass \m\, moving in a center field of a large mass \M.
Below, $m$ without a subscript always refers to the rest mass. The equations of motion in polar coordinates are\\
\be
\label{4a}
\ddot{r}-r\dot{\phi}^2=\frac{1}{f}(-\frac{G M}{r^2}-f'\dot{r}^2)
\ee

\be
\label{4b}
\frac{d}{dt}(m f r^2 \dot{\phi})=\frac{d}{dt}(f l)=0
\ee
where $'$ means the derivative w.r.t. $r$ and $f l = f m r^2 \dot{\phi}$ is the magnitude of the angular momentum.\\

Because the scalar field \f\, doesn't influence the direction of the gravitational force, the angular momentum vector
stays fixed in space. Thus, eq.(\ref{4b}) expresses the conservation of angular momentum.\\

In a circular orbit, $\dot{r}=0$ and $f'=0$, and therefore the circular velocity is
\be
\label{5}
v_C=r \dot{\phi}= \sqrt{\frac{1}{f} \frac{GM}{r}}
\ee
At large distances ($r \gg \lambda$) we use $f \asympt \lambda /r $ to obtain

\be
\label{5a}
v_{C} \asympt \sqrt{\frac{GM}{\lambda}}=\sqrt[4]{G M a_C}=const
\ee
showing that the circular velocity at large distances from the mass \M\, becomes constant in \mmc. MOND yields the same equation with
$a_C=a_M$, where $a_M$ is the MOND parameter\footnote{in MONDian literature denoted as $a_0$ (see below)}.
This is what is observed at the outskirts of galaxies.\\

Unlike the DM paradigm the \mmc\, requires not more gravitational, but less inertial mass.\\

Furthermore, there is a simple explanation for the fact that the virial mass $M_{vir}$ is always larger than the
observed mass $M_{obs}$.\\

The virial theorem can be derived by defining a quantity  $\cal{G}$ $= f \vp \cdot \vr$  \cite{Goldstein}. The
time average for a finite motion  then is $\langle {\cal{G}} \rangle=0$. The derivative of $\cal{G}$ w.r.t. time  and subsequent
averaging together with eq.(\ref{3}) yields  $\langle 2fT \rangle=\langle \nabla U\cdot \vr \rangle$, where T is the usually
defined kinetic energy $T=\frac{1}{2} m v^2$. For the Newtonian gravitational potential,  the resulting virial
theorem in \mmc\, is
\be
\label{6}
2\langle f T \rangle =-\langle U \rangle
\ee

Interestingly there is no expression resulting from the second inertial force. Thus, if
$\langle f v^2 \rangle \approx \langle f  \rangle  \langle v^2 \rangle$
holds, the virial mass according to  \mmc\,  is smaller than the Newtonian virial mass $M_N$ by a factor
$\langle f \rangle$
\be
\label{7}
M_{ \mind} \approx \langle f\rangle M_N
\ee

We now  discuss the asymptotic case  $ r \gg \lambda $ for the second term on the right-hand side of eq.(\ref{4}),
where we have $f \asympt \frac{\lambda}{r}$.

Using $\dot{f}=\nabla f \cdot \vv= f' \hat{r} \cdot \vv$ this term, $ \displaystyle{ -\frac{\dot{f}}{f}\vv}$,
asymptotically becomes

\be
\label{8}
-\frac{\dot{f}}{f}\vv \asympt -\frac{-\frac{\lambda}{r^2}\hat{r}\cdot\vv}{\frac{\lambda}{r}}\vv=\frac{\hat{r}\cdot\vv}{r}\,\, \vv
\ee

which does not depend on  $\lambda$! This means that at large distances the repulsion is independent of mass \M.

Thus, the second term in eq.(\ref{4}) represents nothing else but a non-relativistic analogue to the Hubble flux.
A convincing argument in favor of this interpretation is the fact that in the cosmological context the position
of each object may be written $\vr= a(t)\vec{\xi}$ with the scale factor $a(t)$ and the constant comoving position $\vec{\xi}$.
Then the term \,$\ds -\frac{\dot{f}(|a\vec{\xi}\,|,\lambda)}{f(|a\vec{\xi}\,|,\lambda)}$ \,asymptotically becomes the
Hubble function $\ds H(t)=\frac{\dot{a}}{a}$\,. It depends neither on $\vec{\xi}$ nor on $\lambda$.
However, a correct treatment of the Hubble flux is only possible within the framework of GR and will be discussed in
sections\ref{mmccosmol} and \ref{observ}.

\subsection{Energy Conservation}
Now we turn to the energy conservation theorem for a test mass \m\, whose  potential energy $U$ in the field of a large
mass \M\,is not explicitly time-dependent. The additional inertial force produces an additional term.
Starting from eq.(\ref{3}) and taking the scalar product with $\vv$,
one obtains
\[
f m  \dot{\vv}\cdot \vv +  \dot{f} m \vv \cdot \vv=-\nabla U \cdot \vv=-\frac{dU}{dt}
\]
or
\[
m\frac{d}{dt}(\frac{1}{2} f \vv ^2) + \frac{1}{2} \dot{f} m v^2=-\frac{dU}{dt}
\]
The total energy is now obtained by integrating over t
\be
\label{9}
\frac{1}{2} f m   v ^2 + \frac{1}{2} \int  \dot{f} m v^2 \, dt + U = E
\ee
The meaning of the second term is discussed in section\ref{mmccosmol}.\\

\subsection{\mmc\, vs. MOND}
Today there is a completely relativistic version \cite{Bekenstein05}  of MOND. The difference between MOND and \mmc\,
however can be seen  most clearly by analyzing the non-relativistic equation of motion.
While in \mmc\, eq.(\ref{3}) holds, the corresponding equation in MOND according to \cite{Bekenstein05} is
\be
\label{10}
\mu \left( \frac{|\vvp|}{a_M} \right)\vvp=-\nabla \Phi_N
\ee
where $\mu(x)$ is an empirical function with the following properties\\
\be
\label{11}
\mu(x)=\left\{ \begin{array}{ll} 1 & x\to \infty\\ x & x \ll 1 \end{array} \right.
\ee

and $a_M=0.169 c H_0 = 1.2\, 10^{-10} m/s^2$ is the parameter of MOND.\\

The following function is often used for $\mu(x)$\cite{Sanders96}
\be
\label{12}
\mu(x)=\frac{x}{\sqrt{1+x^2}}
\ee

in the limit $|\vvp| \ll a_M$ eq.(\ref{10}) becomes
\be
\label{10a}
\frac{|\vvp|}{a_M} \vvp=-\nabla \Phi_N
\ee

As Bekenstein\cite{Bekenstein05} pointed out, "MOND is characterized by a scale of acceleration $a_M$, not by a scale of
length". This is a fundamental difference; \mmc\, is characterized by a scale of length $\lambda$.\\

MOND has two interpretations \cite{Milgrom01}:
\benum
\item[a)] as a modification of the Newtonian equation of motion, as given by eq.(\ref{10}), without
          a change of the gravitational field.
\item[b)] as a modification of gravity.\\
          In doing so a generalized Poison equation has to be solved
          \be
          \label{13}
          \nabla \cdot ( \mu(\frac{|\nabla \tilde{\Phi}|}{a_M}) \nabla \tilde{\Phi}) = 4\pi G \rho
          \ee
          where $\rho$ is the mass density.
          Together with the solution of eq.(\ref{13}), the unchanged Newtonian equation of motion holds
          $\vvp=-\nabla \tilde{\Phi}$.
\eenum

In \mmc, the gravitational potential and the structure of the equation of motion remain untouched, merely the definition
of mass has changed.

%version 099  5.12.2010

\section{\mmc\, and Relativistic Mechanics}
\label{mmcrelmech}
Below I use the notations  $i,j,k,\ldots=0,1,2,3$ and  $\alpha=1,2,3$. The time coordinate has the subscript 0
and the signature is + - - -.\\

In order to obtain the law of energy conservation in SR,  we first need a generalization of Newton's second law valid
for all velocities. For a test mass \m\, with the potential energy U which does not explicitly depend on time, the
generalized second law is
\be
\label{15}
\frac{d}{dt} \frac{ f m \vv}{\wurz}=-\nabla U
\ee
Using the same procedure as in the derivation of eq.(\ref{9}),
it is easily found that
\be
\label{16}
f \frac{m c^2}{\wurz} - \int\, m c^2 \, \wurz\, \dot{f} \, dt + U = E
\ee\\

To obtain a covariant\footnote{in terms of invariance under Lorentz transformation} equation of motion, I use
the invariant four-dimensional element of length $ds=\sqrt{dx_i dx^i}$ and the dimensionless 4-velocity
\[
\label{17}
u^i=\frac{dx^i}{ds}=(\frac{1}{\wurz}, \frac{\vv/c}{\wurz})
\]
and the resulting 4-momentum is
\[
\label{18}
f p^i = f m c u^i
\]
A Minkowski force can be defined with the 3-force $\vec{F}= -\nabla U$
\[
\label{19}
K^i=(\frac{\vec{F} \cdot\vv/c}{\wurz}, \frac{\vec{F}}{\wurz})
\]
The desired equation of motion using the proper time interval  $d\tau=\wurz\, dt$ is
\be
\label{20}
\frac{d}{d\tau}(m c u^i)=\frac{1}{f}( K^i - m c w^i \frac{df}{d\tau})
\ee
with the 4-vector
\[
\label{21}
w^i=(\frac{v^2/c^2}{\wurz},\frac{\vv/c}{\wurz})
\]
and for $u^i$ and $w_i$ one obtains
\[ u^i w_i =0 \]

Eq.(\ref{20}) is the analog to eq.(\ref{4}) and can be rewritten with $ds= c d\tau$
\be
\label{20a}
\frac{du^i}{ds}=\frac{1}{f}(\frac{K^i}{m c^2}-w^i\frac{df}{ds})
\ee
the time coordinate of eq.(\ref{20a}) yields the energy conservation eq.(\ref{16}), the space coordinates of
eq.(\ref{20a}) the three dimensional equation of motion (\ref{15}).\\

In GR gravitation is included in the metric  $g^{ik}$. Hence, we set $K^i=0$ in eq.(\ref{20a}) and generalize to
arbitrary coordinates by using the covariant derivative instead of the partial one. Then we find the desired result for GR
\be
\label{20b}
\frac{du^i}{ds} + \Gamma^i_{kl} u^k u^l = - \frac{1}{f} \frac{df}{ds} w^i
\ee
Now $u^i$ and $w^i$ generally depend on $g^{kl}$.  $\Gamma^i_{kl} $ are the Christoffel symbols.\\

%version 099  5.12.2010

\section{\mmc\, and Cosmology}
\label{mmccosmol}

\subsection{The Equation of State}
In this section $a$ is the scale factor with dimension of length.\\

Now I discuss the physics of the second inertial force and its justification.
An adequate model for homogeneous and isotropic matter in cosmology is that of a perfect fluid, whose energy-momentum-tensor
is given by
\be
\label{22}
T^{ik}=(\epsilon+P)u^i u^k- \rho g^{ik}
\ee
P is the pressure, $\epsilon$ the energy density in the proper system.\\
It describes the background and therefore we have to replace \f\, by \fb\, in this context.

If we choose a synchronous reference system, then  $x^0=c\tau$ as well as  $g_{00}=1$ and $g_{0\alpha}=0 $.
Otherwise  $g_{00}$ and  $u^0$ would depend on the gravitational potential $\Phi$. Then
\be
\label{22a}
T^{00}=\frac{\epsilon + P}{\wurz^{\:2}}-P=\frac{\epsilon+P\:\frac{v^2}{c^2} }{\wurzq}
\ee

Since $T^{00}$ is the energy density in the frame of reference, it is determined by the energy conservation law
(\ref{16}). For that purpose we set $U=0$ and obtain
\be
\label{23}
f_b \frac{mc^2}{\wurz}- \int_0^\tau m c^2 \wurz \:\frac{df_b}{d\tau'}\: d\tau' = E
\ee
Now in eq.(\ref{23}) we switch over to densities remembering that the volume has to be transformed too.
\[
f_b \frac{\rho c^2}{\wurzq}-\frac{1}{\wurz} \int_0^\tau \rho c^2 \wurz\: \frac{df_b}{d\tau'}\: d\tau' =\frac{E}{V}
\]
setting
\[
\label{25}
 F(\tau)= \int  \wurz \: \frac{df_b}{d\tau}\: d\tau
\]
we obtain
\[
\rho c^2 \frac{f_b-\frac{F(\tau)-F(0)}{\wurz}+ \frac{v^2}{c^2}\,\frac{F(\tau)-F(0)}{\wurz}}{\wurzq}=\frac{E}{V}
\]

Now we compare this with eq.(\ref{22a}) keeping in mind that there $\epsilon$ and $P$ are quantities in the proper
system, where $ F(\tau)=f_b(\tau)$. If we refer to cosmic time, then $f_b(0)=1$ is valid, because the initial state is
characterized by an extreme density of mass and energy respectively.\\

Thus,
\be
\label{27a}
\epsilon=\rho c^2
\ee
\be
\label{27b}
P=(f_b-1)\rho c^2
\ee
Eq.(\ref{27b}) is the equation of state of matter in \mmc. According to \mmc\, matter  always shows negative pressure
except from $f_b=1$.\\

\subsection{Cosmological Consequences}
In this subsection the subscript $0$ denotes the present epoch.\\
Below, radiation will be neglected and \LCDM\, always refers to the two component \LCDM.\\

In the cosmological context the scalar field \fb\, is considered as a function of the scale factor $a$.
Often the dimensionless state of equation \w\, is used \cite{Sami09}
\be
\label{28}
w=w(a)=\frac{P}{\epsilon}=f_b-1
\ee

While in \mmc\, \w\, is a function of the scale factor $a$,  \w\, is constant in \LCDM. For pressureless matter
\LCDM\, yields $ w_m=0$, and for the cosmological constant $\Lambda$  $w_{\Lambda}=-1$ holds.
In other scalar field models of DE, e.g., the Quintessence \cite{Copeland06} $w=w(\varphi)$ with a scalar field
$\varphi$  holds.\\

The basic equations of cosmology are:\\
the two Friedmann equations (here without $\Lambda$)\\
\be
\label{29}
H^2=(\frac{\dot{a}}{a})^2=\frac{8 \pi G}{3} \rho - \frac{k c^2}{a^2}
\ee
and
\be
\label{30}
\frac{\ddot{a}}{a}=- \frac{4\pi G}{3} (\rho + \frac{3P}{c^2})
\ee
as well as the continuity equation
\be
\label{31}
\dot{\rho}+ 3 H (\rho +\frac{P}{c^2})=0
\ee
and the equation of state(\ref{27b}).

We substitute eq.(\ref{27b}) into eq.(\ref{30}) and obtain
\be
\label{30a}
\frac{\ddot{a}}{a}= -4\pi G \rho ( f_b-\frac{2}{3})
\ee

From this it follows the important result that for $f_b<2/3$ the expansion of the universe is accelerated ($\ddot{a}>0$). This is
what we observe.\\

Substituting eq.(\ref{27b}) into eq.(\ref{31}) we find
\be
\label{31a}
\dot{\rho}+3 H f_b \rho = 0
\ee

From this continuity equation we obtain also some interesting results.\\

The solution of eq.(\ref{31a}) is
\be
\label{32}
\rho(a)=\rho_0 e^{\,-3 \, \int_{a_0}^a \frac{f_b(a')}{a'}\, da'}
\ee
The scalar field may be written as $f_b(a)=1-\phi(a)$\footnote{$\phi$ is quasi a measure for the deviation from standard theory}.
Thus,
\be
\label{32a}
\rho(a)=\rho_0 (\frac{a_0}{a})^3 \, e^{\,3\, \int_{a_0}^a \frac{\phi(a')}{a'}\, da'}
\ee
and defining the enhancement factor
\be
\label{33}
\Gamma(a)= e^{3\,\int_0^a \frac{\phi(a')}{a'}\, da'}
\ee
we obtain
\be
\label{32b}
\rho(a)=\rho_0 (\frac{a_0}{a})^3 \,\frac{\Gamma(a)}{\Gamma(a_0)}
\ee

The attenuation effect of the pressureless matter  $\propto 1/a^3$ during the expansion is separated as a factor.
Since $f_b(a)$ asymptotically approaches zero, $\phi(a)$ trends towards 1.
Therefore, according to eq.(\ref{32a}),  $\rho$ approaches a positive limit \rhoinf. Thus, the universe  finally trends to
a constant energy density.\\

To distinguish analog quantities in \mmc\, from their counterpart in \LCDM\, the former will be denoted with  $\tilde{ \,}$.

It is intuitively obvious that we can define a cosmological constant in \mmc\, in this way:
\be
\label{35}
\rhoinf=\tilde{\rho}_{\Lambda}=\frac{c^2}{8 \pi G}\tilde{\Lambda}
\ee

Thus, the cosmological scale length $\lambda$ is also found
\be
\label{36}
\lambda=\frac{1}{\sqrt{\tilde{\Lambda}}}=\frac{\sqrt{3} c}{H_{\infty}}
\ee

In the cosmological context $\lambda$ depends on the asymptotic density of mass and hence on the asymptotic Hubble constant
 $H_{\infty}$.\\

Now we will divide the density $\rho$ in the two fictitious (!) parts  $\tilde{\rho}_m$ and $\tilde{\rho}_{DE}$, which
in \LCDM\, correspond to the pressureless (baryonic as well as Dark) matter and the DE in terms of the cosmological constant.
But, in \mmc\,  $\tilde{\rho}_{DE}$ isn't constant.\\

Once again we have to keep in mind that in \mmc\, there is neither DM nor DE!\\

With
\be
\label{Y1}
\rho=\tilde{\rho}_m + \tilde{\rho}_{DE}
\ee
 and
\be
\label{Y2}
 \tilde{\rho}_m=\tilde{\rho}_{m,0}(\frac{a_0}{a})^3
\ee
we find
\be
\label{KS3}
\tilde{\rho}_{DE}= \rho_0 (\frac{a_0}{a})^3 \left( \frac{\Gamma(a)}{\Gamma(a_0)}- \frac{\tilde{\rho}_{m,0}}{\rho_0}   \right)
\ee
Since  pressure as well as $\tilde{\rho}_{DE}$ vanish if $a \ll \lambda$ and using  $\Gamma(0)=1$ we find
\be
\label{KS4}
\Gamma(a_0)= \frac{\rho_0}{\tilde{\rho}_{m,0}}
\ee

Thus, we can rewrite eq.(\ref{32b}) as
\be
\label{X1}
\rho =\tilde{ \rho}_m \, \Gamma(a)
\ee

Assuming that  $\rho$ is always critical and with the scaled density eq.(\ref{X1}) becomes
\be
\label{X2}
 \tilde{\Omega}_m \, \Gamma(a)=1
\ee
 or,
\be
\label{X3}
\tilde{\Omega}_{DE}= \tilde{\Omega}_m (\Gamma(a) - 1)
\ee

Since eq.(\ref{31a}) has constant solutions for  $f_b=0$, a simple possibility opens up to explain the occurrence
of the inflation.

Let us assume that at Planckian time the scalar field \fb\,  has the value zero and the inflation began at  $t^*$.
If the value of the scalar field switches to 1 at a later time  $t_i$ by symmetry breaking, then this results in a very
high constant density $\rho^*$ during $t^* \ge t \ge t_i$. According to eq.(\ref{27b}) a very high negative pressure occurs
that fuels the inflation. At time  $t_i$ the inflation stops abruptly.\\

So in \mmc\, the initial and the final state of the universe is characterized by constant densities and thus by
massless matter (de Sitter universe). The well-known de Sitter solution during inflation is
\be
\label{38a}
a(t)=a^* e^{H^* (t-t^*)} \mbox{ \quad with \quad} H^*=\sqrt{\frac{8 \pi G}{3}\, \rho^*}
\ee
and in the final state
\be
\label{38b}
a(t)=a_0 e^{H_{\infty}(t-t_0)}  \mbox{ \quad with \quad} H_{\infty}=\sqrt{\frac{8 \pi G}{3} \rhoinf}
\ee\\

The flatness condition  of space ($k=0$) follows from of the Friedmann eq.(\ref{29})
\begin{eqnarray}
\label{39}
k & = & \frac{a^2}{c^2}(\frac{8 \pi G}{3}\rho - H^2) \nonumber\\
  & = & \frac{8 \pi G}{3\, c^2}\rho_0\, \frac{a_0^3}{\Gamma(a_0)} \frac{\Gamma(a)}{a}-\frac{\dot{a}^2}{c^2}=0
\end{eqnarray}
using
\[
\dot{a}^2=\frac{8 \pi G}{3}\rho_0 \frac{a_0^3}{\Gamma(a_0)} \frac{\Gamma(a)}{a}=
H_0^2 \frac{a_0^3}{\Gamma(a_0)} \frac{\Gamma(a)}{a}
\]
we obtain
\be
\label{39a}
\int_0^a \frac{\sqrt{a'}\, da'}{\Gamma(a')^{1/2}}=H_0\frac{a_0^{3/2}}{\Gamma(a_0)^{1/2}} t
\ee

In section \ref{observ} it is shown that eq.(\ref{39a}) delivers the correct age of the universe.

% version 099  5.12.2010

\section{\mmc\, and Observations}
\label{observ}

Since up to now there is no field function \f\, derived from first principles, two empirical functions will be tested for
their ability to reproduce various observations.
In the cosmological context as background field function, I used
\[
f_1(x)  =  1-e^{-1/x}\mbox{\qquad with \qquad}  x=\frac{a}{\lambda}
\]

For the galactical rotation curves, the function
\[
f_2(x) = \frac{1}{\sqrt{1+x^2}}\mbox{\quad with \quad}  x=\frac{r}{\lambda}
\]
is used. The function $f_2$ is related to the empirical function of MOND (eq.(\ref{12}))
by $f_2(x)=\mu(1/x)$.\\

\subsection{Galactical Rotation Curves}

By means of two arbitrarily selected examples, the NGC 5033 and the low surface-brightness galaxy UGC 128, it is shown
that MOND and \mmc\, yield very similar results. The measurements used are from \cite{Sanders96}.\\

The method chosen here is geared to the approach of MOND in \cite{Sanders96}.
There, at first the enclosed total mass $M_t(r)$ of a circular orbit with radius $r$ is divided into the disk mass
$M_d$, the gas mass $M_g$ and bulge mass $M_b$. Thus $M_t=M_d+M_g+M_b$. The examples are chosen so that $M_b=0$
to minimize the uncertainties. The partitioning was made under the assumption that Newtonian mechanics is valid.
Hence,
\[
M_t(r)=\frac{(v_d^2+v_g^2)r}{G}
\]
Now the MOND eq.(\ref{10}) becomes
\be
\label{rot1}
\mu \left( \frac{\dot{v}}{a_M} \right) \dot{v}=\frac{G M_t(r)}{r^2}
\ee

It has to be solved for the acceleration $\dot{v}$. Then the rotation curve velocity is given by
\be
\label{rot2}
v_{MOND}(r)=\sqrt{r \dot{v}(r)}
\ee
In \mmc\, instead of eq.(\ref{rot1}) one has to solve the following eq.
\be
\label{rot3}
\frac{1}{f(r)}\, \frac{G M_t(r)}{r^2}= \frac{v^2}{r}
\ee

So we find
\be
\label{rot4}
v_{MMC}=\sqrt{\frac{1}{f(r)}(v_d^2+v_g^2) }
\ee
The scale length $\lambda$ is given by
\[
\lambda(r)=\sqrt{\frac{G M_t(r)}{a_{C}}}
\]
with the fit parameter $\alpha$ defined by $a_{C}=\alpha c H_0$.\\

\begin{figure}[H]
 \centering\includegraphics[width=10cm,height=12cm, angle=-90]{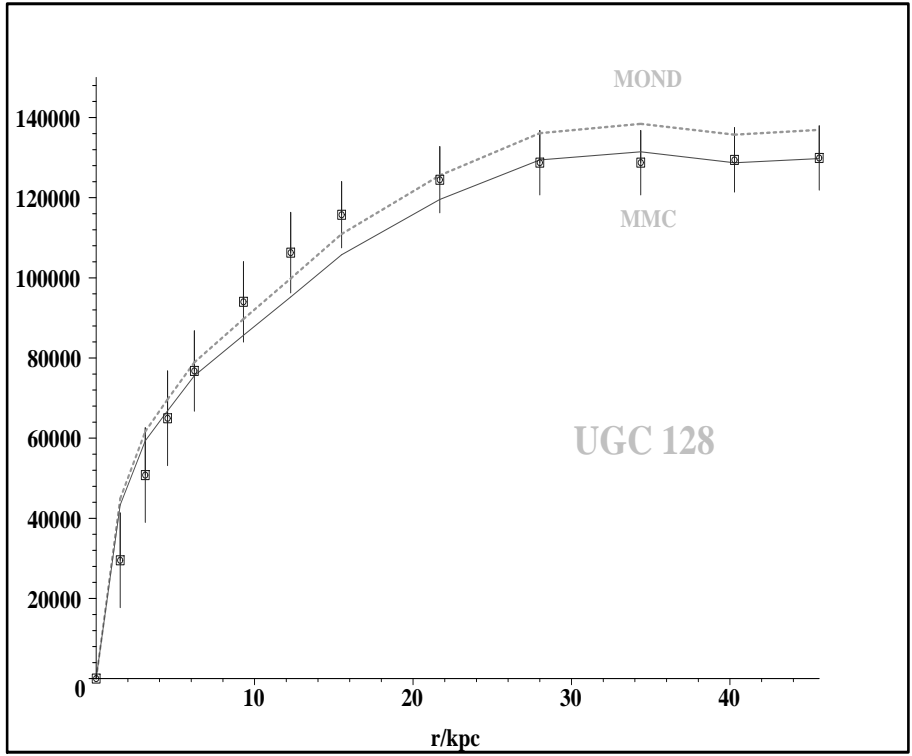}
 \caption{ \small \label{UGC128} Rotation curve of UGC 128. Velocity in m/s.
         \mmc\, solid line; MOND dashdot line.
         Data points from  \protect\cite{Sanders96}  }
\end{figure}

\begin{figure}[H]
 \centering\includegraphics[width=10cm,height=12cm, angle=-90]{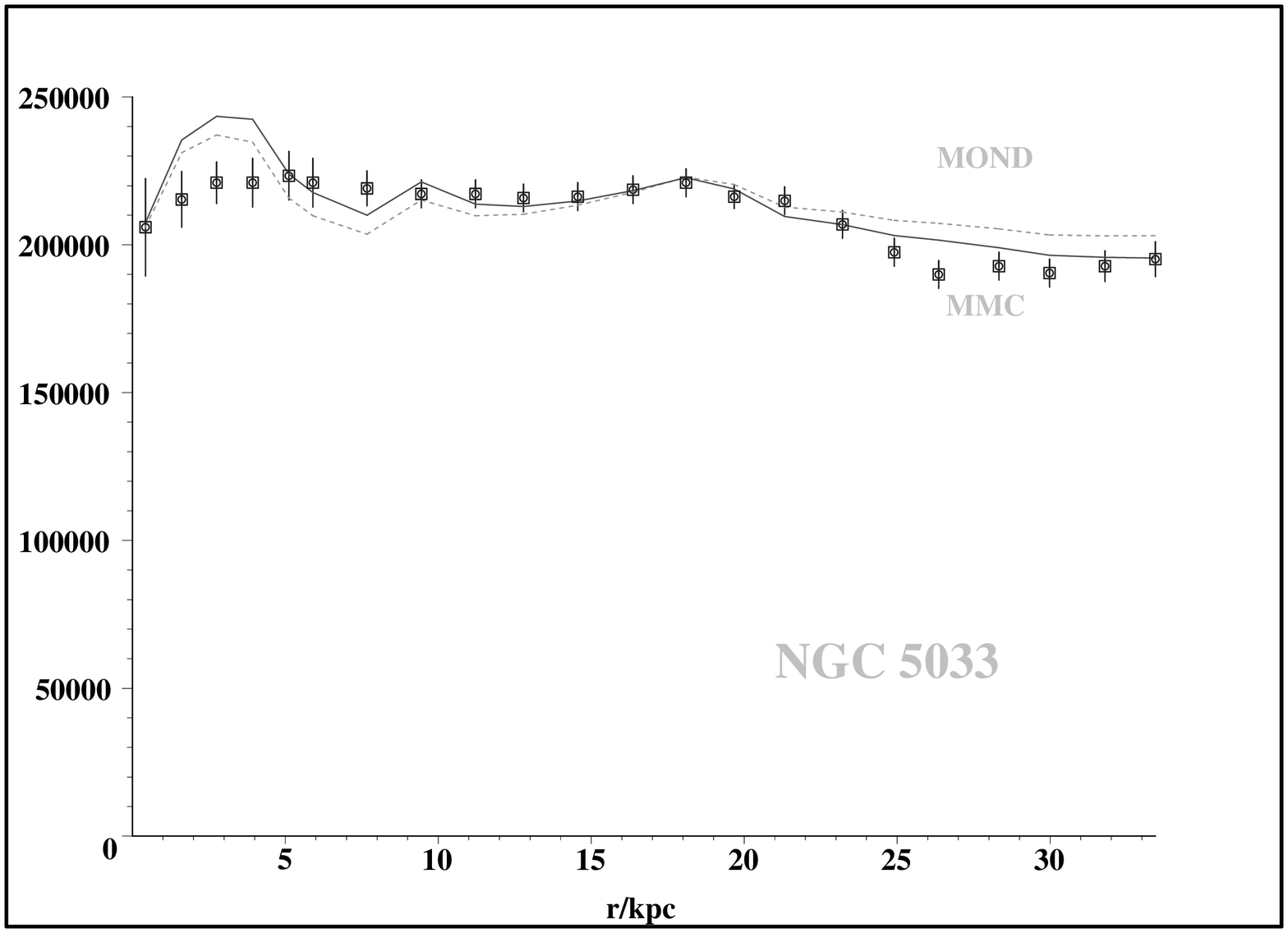}
 \caption{ \small \label{NGC5033} Rotation curve of NGC 5033. Velocity in m/s.
         \mmc\, solid line; MOND dashdot line.
         Data points from  \protect\cite{Sanders96}   }
\end{figure}

The rotation curves fig.\ref{UGC128} and fig.\ref{NGC5033} are computed with the empirical function $f_2$ also used in MOND. The results are very insensitive to
the chosen function. The function $f_1$ delivers almost the same results with a somewhat different fit parameter.
For $f_1$ we find $\alpha=0.101$ and for $f_2$  $\alpha=0.135$;
for comparison the MOND fit was performed with $a_M=0.169cH_0$.

Both examples suggest that whenever MOND works \mmc\, will  work too.\\

\subsection{Cosmological Aspects}
The cosmological \mmc\, model needs 3 input parameters:
$\tilde{\Omega}_{m,0}$, the normalized density of the pressureless matter of the present epoch,
$H_0$ the present Hubble constant and $k$, the  curvature parameter of the metric.\\
The values used here are: $\tilde{\Omega}_{m,0}=0.25$, $H_0=0.237\cdot 10^{-17}/s$ and $k=0$.

The flatness of space is presumed not explained. \mmc\, is a "one component" model. However, because space is flat this
single component must be critical.\\
Hence a series of predictions arise  agreeing more or less with the observations.\\

The following computations are performed with function $f_1$. Analogous expressions for $f_2$ are
easily found. The results are not strongly dependent on the chosen function.

It should be remembered that the functions used are merely empirical. Therefore, we should not expect a perfect correlation
with observation.\\

From $f_1=1-e^{-\lambda/a}$ we find the enhancement factor\footnote{For the function  $Ei(1,z)$ see \cite{Abram},
there denoted as $E_1(z)$.}

\be
\label{KS1}
\Gamma(a)=e^{3 Ei(1, \lambda/a)}
\ee

We can rewrite eq.(\ref{X2}) for the present epoch
\be
\label{KS2}
\tilde{\Omega}_{m,0}\,  e^{3\, Ei(1, \lambda/a_0)} = 1
\ee
Immediately there result some numerical values:\\
$\Gamma(a_0)=4.00$, the ratio $x_0=a_0/\lambda=1.690$  and $f_1(x_0)=0.447$.\\

Substituting $f_b=f_1$ in eq.(\ref{32b}) we find for $a \to \infty$
\be
\label{KS5}
\rho_{\infty}=\rho_0 \left( \frac{a_0 e^{-\gamma}}{\lambda e^{ Ei(1, \lambda/a_0)}} \right)^3
\ee

where $\gamma=0.5772\ldots$ is the Eulerian constant.\\

Therefore, $\rho_{\infty}=0.214 \rho_0$ and $H_{\infty}=0.462 H_0$.\\
Setting  $\rho_{c,0}=1.01\cdot 10^{-26} \,\,kg/m^3$  we find with eq.(\ref{35}) and eq.(\ref{36})\\
$\tilde{\Lambda}=0.403\cdot 10^{-52}\,\, m^{-2}$,
$\lambda=0.158\cdot 10^{27} m$ and $a_0=0.266\cdot 10^{27} m$.\\

Comparing with \LCDM\,($\Omega_{\Lambda,0}=0.7$), it is clear that the cosmological constant
$\Lambda=1.32\cdot 10^{-52}\,\, m^{-2}$ is slightly larger than $\tilde{\Lambda}$.\\

Substitution of the values above into eq.(\ref{39a}) delivers the age of the universe as $t_0=12.9 Gy$, which
has the right order of magnitude.\\

The two component \LCDM and \mmc\, mostly produce similar results. This shall be shown with the help of 3 examples.
\benum
\item[$\alpha$)]
The normalized density of the cosmological constant in \LCDM is (see \cite{Copeland06})
\be
\label{KS9}
\Omega_{\Lambda}=\frac{\Omega_{\Lambda,0}}{\Omega_{m,0}(\frac{a_0}{a})^3+\Omega_{\Lambda,0}}
\ee

\begin{figure}[H]
 \centering\includegraphics[width=10cm,height=12cm, angle=-90]{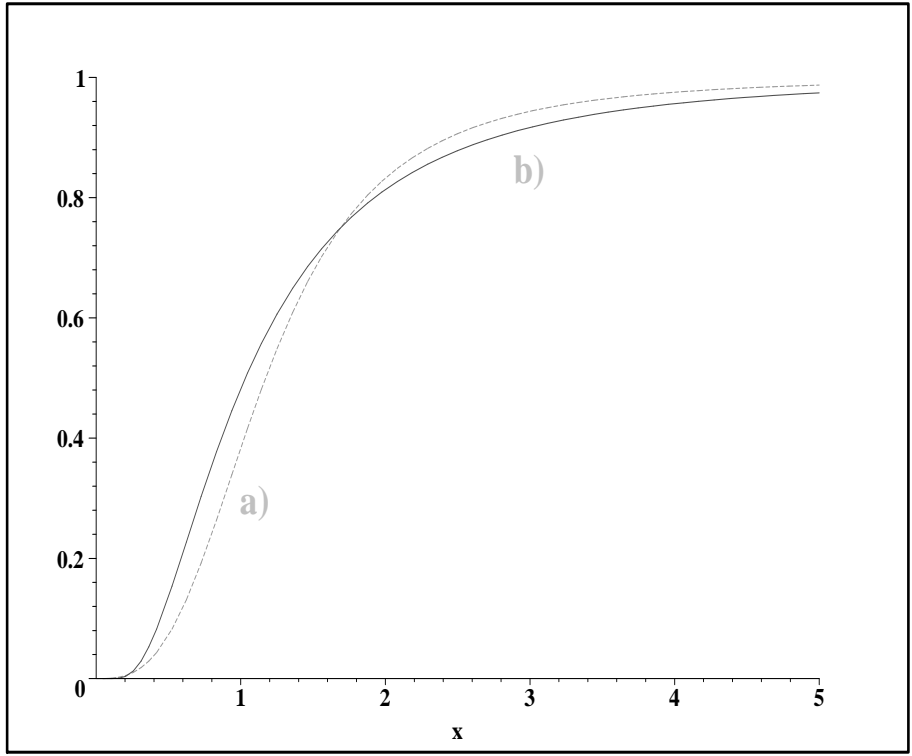}
 \caption{ \small \label{OmDE}  a) $\Omega_{\Lambda}$ and b) $\tilde{\Omega}_{DE}$  as a function of $x=a/\lambda$}
\end{figure}

With eq.(\ref{X3}) \mmc\, yields
\be
\label{KS10}
\tilde{\Omega}_{DE}=\tilde{\Omega}_{m,0} (\frac{a_0}{a} )^3 ( e^{ 3 Ei(1,\lambda/a)} - 1)
\ee
Both functions are shown in fig.\ref{OmDE}. The curves cross over at $x_0$ if $\tilde{\Omega}_{m,0}=\Omega_{m,0}$.
\item[$\beta$)]
The luminosity distance $d_L$ in \LCDM with the redshift parameter z
\be
\label{redshift}
\frac{a_0}{a} = 1+ z
\ee
is according to \cite{Copeland06}
\be
\label{KS11}
d_L = \frac{1+z}{H_0} \, \int_0^z \frac{dz'}{\sqrt{\Omega_{m,0}(1+z')^3 + \Omega_{\Lambda,0}}}
\ee
In \mmc\, we find with eq.(\ref{KS2})
\begin{eqnarray}
\label{KS12}
\lefteqn{d_L =}\\
&&\frac{1+z}{H_0} \int_0^z \frac{dz'}{\sqrt{ (1+z')^3 e^ {3(Ei(1,\frac{1+z'}{x_0})-Ei(1,\frac{1}{x_0}))} }}\nonumber
\end{eqnarray}

\begin{figure}[H]
 \centering\includegraphics[width=10cm,height=12cm, angle=-90]{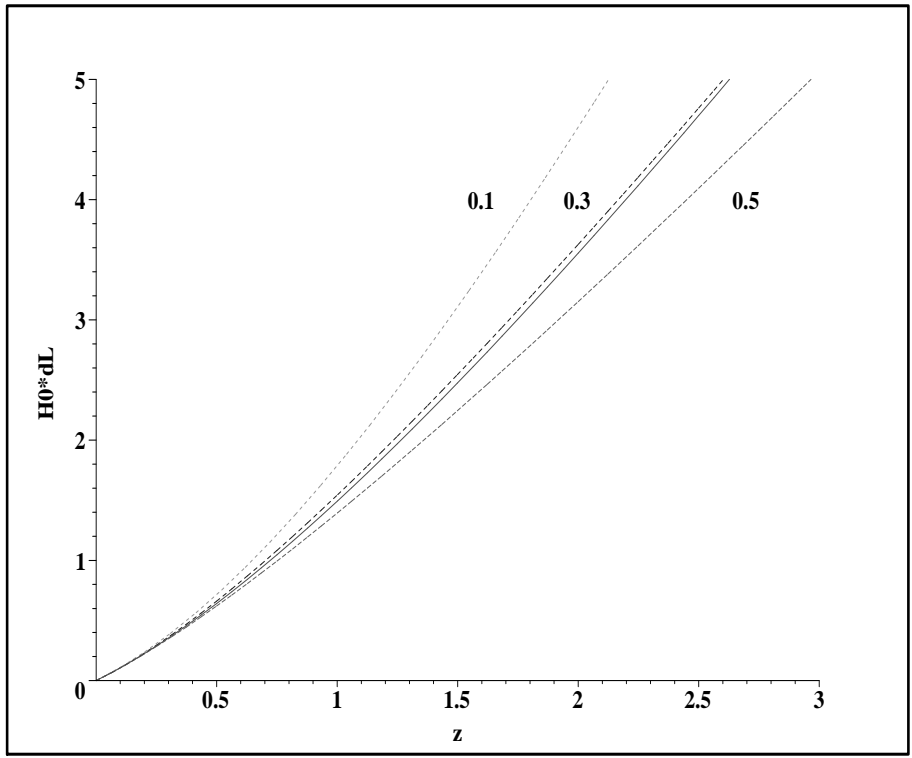}
 \caption{\label{H0dL} \small  $H_0\cdot d_L$ vs. $z$. Solid line is \mmc. The parameters are  $\Omega_{m,0}$ for \LCDM }
\end{figure}

For a comparison, see fig.\ref{H0dL}

\item[$\gamma$)]
In \LCDM the accelerated expansion of the universe began at \cite{Copeland06}
\[
z_c = (\frac{2 \Omega_{\Lambda,0}}{\Omega_{m,0}})^{1/3}-1
\]
With $\Omega_{\Lambda,0}=0.7$, this gives $z_c=0.67$. Substituting  $\tilde{\Omega}_{\Lambda,0}=0.75$ we find $z_c=0.82$.\\
In \mmc\, we have to solve $f_1(x_{2/3})=2/3$ and find $x_{2/3}=0.910$. This together with eq.(\ref{redshift}) gives
$z_{2/3}=0.86$
\eenum

A substantial difference between standard physics and \mmc\, is that according to the former the peculiar motion
of a body in a complete homogeneous and isotropic expanding universe will finally come to rest in the Hubble flux, while
in \mmc\, this is not the case.\\

In standard physics, a body with momentum $p$ w.r.t. a comoving system (see textbooks, e.g. \cite{LandauL2})
\[
p\, a = \frac{m v a}{\wurz}=const
\]
hence the velocity $v$ decreases while $a$ increases. In \mmc\, in contrast, the momentum is $ f p$, thus
\[
f p \,a =\frac{f m v a}{\wurz}=const
\]

If $f \propto \lambda/a$ holds then $v$ is constant. This could help to understand an observation reported by Watkins et
al. \cite{Watkins08}. They found  that there is an exceptional large bulk flow\footnote{dipole moment of the peculiar
velocity field} of 407 km/s within a Gaussian window of radius $50 h^{-1}$ Mpc. According to the \LCDM\, cosmology the
value should be about 100 km/s, which is significantly too small. Certainly inhomogeneities will have a large influence
on this phenomenon. However, \mmc\, could at least partly explain the observed large value.\\

\onecolumn


%version 099 5.12.2010

\section{Summary}

Some consequences of the assumption that the inertial mass depends not only on its rest energy but also on a
scalar field which is caused by other mass are discussed. Applying this assumption to the equation of motion we find that
besides the well-known inertial force, $-m \vvp$, a further inertial force emerges. This force is always repulsive and favors
the decrease of inertial mass. The acceleration caused by this force becomes independent of mass at large distances.
Thus, it is related to the Hubble flux.

A reduction of inertia on the other hand acts as an enhancement of gravity as if there would be more gravitational mass.
This effect explains the galactical rotation curves. The assumption of a DM isn't necessary
at all. Unlike the DM paradigm the \mmc\, postulates not more gravitational but less inertial mass.\\
It is easy to imagine that the  enhanced gravity will strongly effect the formation of structures at large length scales.
It effects the formation of spiral arms of galaxies just as well as the development of voids and filaments at the
level of super clusters.\\

A new equation of state for baryonic matter has some intriguing properties. For instance, baryonic matter is  always
accompanied by a negative pressure. Therewith, it is a consequence of the Friedmann equation that for  $f_b<2/3$ the
expansion of the universe is accelerated. Moreover, the continuity equation allows constant solutions for $f_b=0$. Thus,
the late universe approaches a state of constant energy density. In this regard, the \mmc\, is similar to \LCDM. It also
opens up the possibility to understand the inflation as a phase transition with $f_b$ as the order parameter. We assume that
immediately after Planckian time the scalar field has the value zero. During this period inflation occurs. Later on
when the value of $f_b$ switches to $1$ by symmetry breaking the inflation stops. Thus, the initial and the final
state of the universe can be understood in a very similar way: it is a state without inertia.\\

Using the empirical functions the computed age of the universe, the luminosity distance and the redshift $z$ of the
beginning of the accelerated expansion comply to large extent with the values of the two component \LCDM with
$\Omega_{\Lambda}=0.7\ldots 0.75$. Therefore, also  DE isn't necessary as explanation.
Furthermore, the \mmc\, can provide an explanation for the very high peculiar velocities found at large scales.\\

From the fact that  \mmc\, requires neither DM nor DE to explain some phenomenons it is incorrect to conclude that these
don't exist. Maybe they do. If \mmc\, is correct, then their energy density should be much smaller as assumed now.\\

\mmc\, is just an idea, not yet a theory.  A lot of questions are still open. The modest aim of this article was
to demonstrate the potential of the idea.\\

%%%%%%%%%%% Ende Version 1 E 13.11.2010

%%%%%\input{xtest}
I am grateful to Richard Davidson for very helpful comments on this paper.\\

\end{document}